\documentclass[acmtog]{acmart}

\usepackage{balance}
\usepackage{colortbl}
\usepackage[utf8]{inputenc}
\usepackage{booktabs} 
\usepackage{pdfpages}
\usepackage{fancyhdr,graphicx,amsmath,amssymb}
\usepackage[ruled,vlined]{algorithm2e}
\usepackage{array}
\usepackage{multirow}
\usepackage{textcomp}
\usepackage{caption}

\PassOptionsToPackage{table}{xcolor}

\setcopyright{none}
\setcitestyle{square}

\renewcommand\footnotetextcopyrightpermission[1]{} 

\begin{document}

\title{A Flexible Neural Renderer for Material Visualization}

\author{Aakash KT}
\affiliation{%
  \department{CVIT, KCIS}
  \institution{IIIT Hyderabad}}
\email{aakash.kt@research.iiit.ac.in}

\author{Parikshit Sakurikar}
\affiliation{%
  \department{CVIT, KCIS}
  \institution{IIIT Hyderabad / DreamVu Inc.}}
\email{parikshit.sakurikar@research.iiit.ac.in}

\author{Saurabh Saini}
\affiliation{%
  \department{CVIT, KCIS}
  \institution{IIIT Hyderabad}}
\email{saurabh.saini@research.iiit.ac.in}

\author{P J Narayanan}
\affiliation{%
  \department{CVIT, KCIS}
  \institution{IIIT Hyderabad}}
\email{pjn@iiit.ac.in}

\renewcommand{\shortauthors}{Aakash KT et al.}

\begin{abstract}
Photo realism in computer generated imagery is crucially dependent on how well an artist is able to recreate real-world materials in the scene. The workflow for material modeling and editing typically involves manual tweaking of material parameters and uses a standard path tracing engine for visual feedback. A lot of time may be spent in iterative selection and rendering of materials at an appropriate quality. In this work, we propose a convolutional neural network based workflow which quickly generates high-quality ray traced material visualizations on a shaderball. Our novel architecture allows for control over environment lighting and assists material selection along with the ability to render spatially-varying materials. Additionally, our network enables control over environment lighting which gives an artist more freedom and provides better visualization of the rendered material. Comparison with state-of-the-art denoising and neural rendering techniques suggests that our neural renderer performs faster and better. We provide a interactive visualization tool and release our training dataset to foster further research in this area.
\end{abstract}



\begin{teaserfigure}
  \centering
  \includegraphics[width=\textwidth]{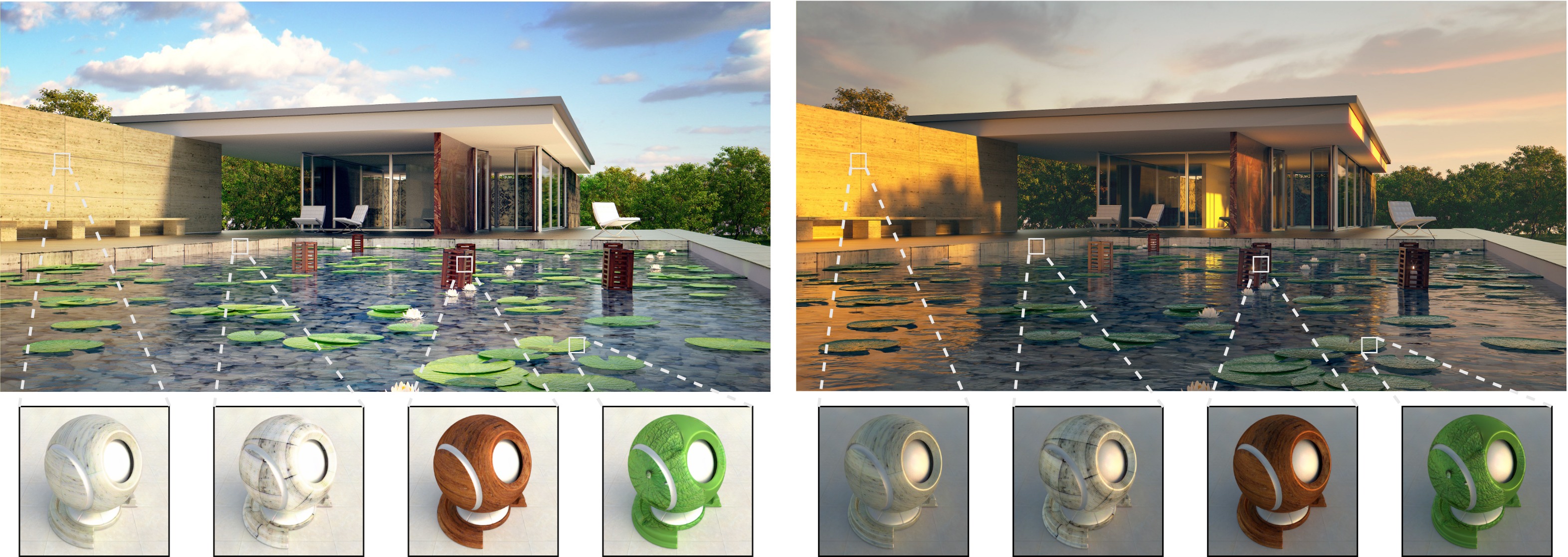}
  \caption{Shaderball visualizations of four selected materials produced by our network are shown at the bottom, for two lighting conditions : \textit{Left} - Daylight and \textit{Right} - Sunset. Scene adopted from \textcopyright eMirage (https://www.emirage.org/)}
  \label{fig:teaser_figure}
\end{teaserfigure}

\maketitle

\thispagestyle{empty}

\section{Introduction}
Ray tracing has emerged as the industry standard for creating photo realistic images and visual effects \cite{Keller:2015:PTR:2776880.2792699}. Accurate modeling of the behavior and traversal of light, along with physically accurate material modeling, creates the beautiful computer generated imagery we see today. Achieving photo realism in such renderings is however a tedious process. Ray tracing is a computationally expensive operation while physically accurate material modeling requires expertise in fine-tuning of parameters to achieve the desired look. Visualization of edits during fine-tuning is very time consuming if the target image is ray-traced. An artist might thereby end up spending a lot of time in a slow and iterative visualization loop.  

In this paper, we present a neural network architecture that can quickly output a high-quality ray-traced visualization of a material. Our work extends the state-of-the-art in material rendering by providing the ability to deal with a large range of uniform as well as spatially-varying materials, along with control over the environment lighting. We render on a fixed shaderball geometry which is complex enough to encode fine interactions between light and the underlying material. 

We evaluate our method quantitatively and also compare qualitative render quality with existing neural rendering frameworks. We also conduct a user study to show the benefit of providing control over lighting for material selection. We show that our proposed system is fast and therefore helps in real-time visualization of materials. Our method also compares favourably with denoising frameworks in producing faster and better rendered images. In summary, the following are the contributions of our work:
\begin{itemize}
    \item A neural renderer to aid in visualization of uniform and spatially-varying materials.
    \item An architectural enhancement to provide control over the environment lighting, thereby increasing visualization capability and freedom.
    \item An interactive tool for material visualization and editing and a large-scale dataset of uniform and spatially-varying material parameters with corresponding ground truth ray-traced images (Project Page\footnote{\url{https://aakashkt.github.io/neural-renderer-material-visualization.html}})
\end{itemize}
Figure \ref{fig:teaser_figure} is an example of the utility of our proposed system. Each material visualization is created within \textit{3 milliseconds} and accurately mimics the behaviour of being rendered in a scene environment.

\section{Related Work}
The use of neural networks for rendering has gained popularity in the recent past. Existing approaches to neural rendering deal with denoising low sample-count Monte-Carlo renders or neural rendering of materials on a fixed geometry. We briefly discuss recent works dealing with denoising, material modelling and acquisition and image-based relighting, in the context of our contributions. 

\textbf{Material modelling:} Several material models have been proposed in the past \cite{Guarnera:2016:BRA:3059330.3059335}. While some models focus on physical accuracy others focus on intuitiveness and simplicity. An intuitive but not strictly physically-accurate material model was proposed by \citet{Burley2012PhysicallyBasedSA}. The model, known as the principled shader model, is parameterized by 13 different values, that control a specific physical aspect of the material. The Cook-Torrance model \cite{Cook:1982:RMC:357290.357293} is another popular material model that is parameterized by four values - \textit{Diffuse Color}, \textit{Specular Color}, \textit{Roughness} and \textit{Normal}. Material model parameters can also be spatially-varying i.e. they can consist of different values at different locations of a 3D surface. In this case, each value is assigned from a \textit{UV-mapping} of the 3D object to a 2D per-pixel parameter map. In our work, we use the Cook-Torrance material model (Section 3).

\textbf{Material acquisition:} Material recovery from a set of images or a single-image is an interesting problem and has been recently studied using neural networks.
A method for accurate capture of BRDF (Bi-directional Reflectance Distribution Function) using a light-stage setup and a deep neural network was proposed by \citet{Kang:2018:ERC:3197517.3201279}. Their approach captures multiple images for accurate reconstruction, where the weights of their network control the illumination within the lightstage. A neural network architecture for recovering the SVBRDF (Spatially Varying BRDF) from a single flash photograph of a planar material was proposed by \citet{DADDB18}. An even more challenging task of recovering the SVBRDF and shape of a free-form object from a single flash photograph is demonstrated by \citet{Li:2018:LRS:3272127.3275055}. While these approaches have significantly improved the speed and accuracy of acquiring materials, the visualization of acquired materials still requires ray-tracing. With advancements in the acquisition process, similar advancements in visualization are imperative, so as to make the whole pipeline function faster. This is the main focus of our work.

\begin{figure*}
    \centering
    \includegraphics[width=\textwidth]{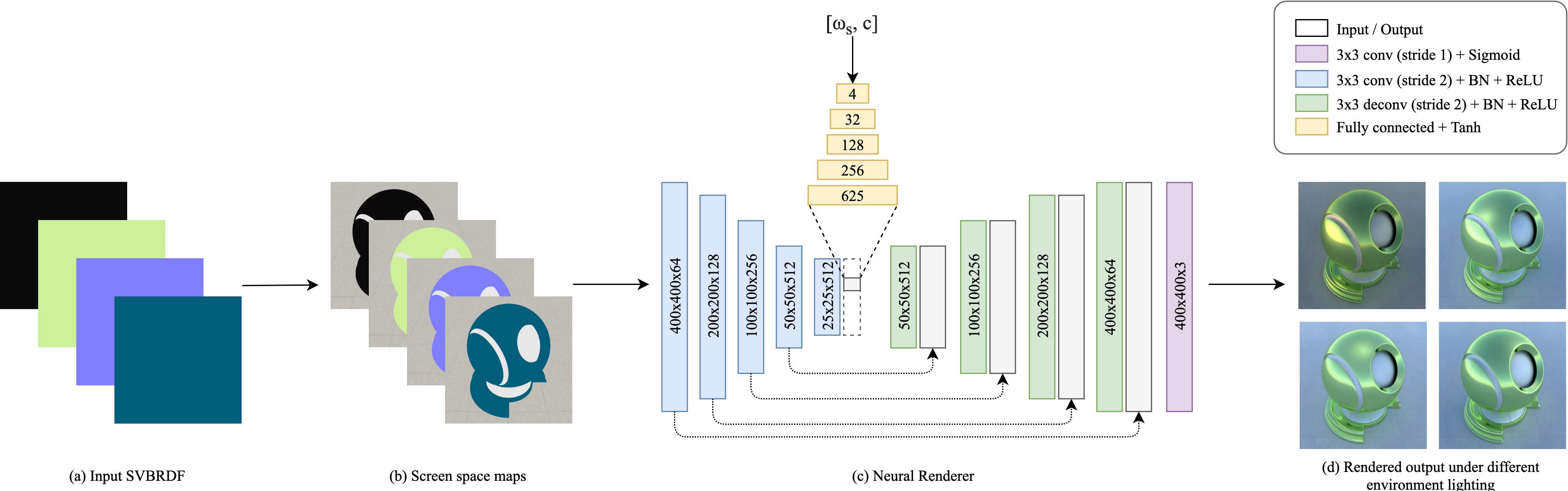}
    \caption{An overview of our proposed workflow. From input SVBRDF maps (a), we create screen-space maps (b) by UV-mapping each map on the shaderball, and rasterizing the scene with that map as base texture. We provide the sun direction and turbidity [$\omega_s, c$] as an input along with the concatenation of screen-space maps. (c) shows the architecture of our proposed neural renderer. (d) shows the results of our network under different environment lighting.}
    \label{fig:architecture}
\end{figure*}

\textbf{Rendering as Denoising:} Much attention has been paid in recent days towards denoising Monte-Carlo (MC) renders in order to enable real-time ray-tracing. A recurrent autoencoder for the task of denoising MC renders was proposed by \citet{Chaitanya:2017:IRM:3072959.3073601} . A more recent method \cite{DBLP:journals/corr/abs-1803-04189} showcases a general denoising framework, trained on noisy input and noisy ground-truth pairs. Both of these approaches make use of auxiliary buffers as additional input to their networks. Other approaches like \cite{Kuznetsov2018DeepAS} aim to efficiently distribute MC samples to bring down the overall render time. All these approaches require a low sample count (spp) MC render as input. One could argue that such methods can be used for quick material visualization, by rendering the a given material with a low sample count, and then denoising the image. We show that a neural renderer specifically designed for material visualization produces faster and better quality results than the corresponding denoising approach.

\textbf{Image-based relighting:} Neural networks have also been used for relighting an image \cite{Ren:2015:IBR:2809654.2766899}. A network that can relight a scene from five differently lit images of the scene was proposed by \citet{xu2018deep}. Yet another neural network for relighting human faces from a single input photograph was proposed by \citet{Sun2019SingleIP}. While not directly related to our work, we take inspiration from such architectures to provide a control over the light direction in the rendered material output. 

\textbf{Neural rendering:} Closely related to our work, neural rendering for material visualization has been proposed by \citet{DBLP:journals/tog/Zsolnai-FeherWW18}. Their approach transforms a low dimensional parameter space to a high dimensional image output, which is the rendered material on a fixed shaderball, under fixed lighting conditions. However, their neural renderer has a large number of parameters, which affects run-time and portability. Also, the target material is rendered under a fixed light position thereby reducing the flexibility of visualization. Additionally, their work only deals with constant material parameters while it is quite common and in fact necessary to have spatially-varying parameters to increase photo realism in materials. Our neural renderer addresses all of these issues, in that it is much smaller and hence faster, allows for control over the light direction and enables the use spatially-varying parameters. We also demonstrate quantitative improvements in rendering quality on comparing our performance with \cite{DBLP:journals/tog/Zsolnai-FeherWW18}.

Ours is a flexible neural renderer for material visualization and can be used in conjunction with material suggestion systems similar to those shown in \cite{DBLP:journals/tog/Zsolnai-FeherWW18}, for building high-quality assistive tools for artists.

\section{Method}
We seek to quickly and accurately render constant or spatially varying material parameters on fixed geometry under controllable environment lighting. The incoming radiance at each pixel \textbf{x} of a such a rendered 2D image can be modeled as:
\begin{equation}
    I(x) = \int_{\Omega} f_{r}(p_{x}, \omega_{i}, \omega_{x}) L(p_{x}, \omega_{i}) (\omega_{i} \cdot n) d\omega_{i},
\end{equation}
where $p_{x}$ is the 3D point corresponding to the 2D pixel \textbf{x}, $\omega_{i}$ is the incoming light direction, $\omega_{x}$ is the direction towards pixel \textbf{x} from point $p_{x}$, $L(p_{x}, \omega_{i})$ is the radiance of incoming light at point $p_{x}$ from direction $\omega_{i}$, $\Omega$ is the set of directions on the upper hemisphere and $f_{r}$ is the Bi-directional Reflectance Distribution Function (BRDF). The choice of $f_{r}$ determines the material model in use. The hyperparameters of $f_{r}$ describe the surface and material properties of the shaderball, which we refer to as the \textit{material parameters}. We refer the reader to \cite{McAuley:2012:PPS:2343483.2343493} for a complete overview of such physically based shading models and \cite{Hoffman2012BackgroundP} for the mathematical details.

We use the Cook-Torrance \cite{Cook:1982:RMC:357290.357293} material model ($f_{r}$) for rendering. Our choice of this material model was based on two aspects: (1) The Cook-Torrance model is based on the microfacet theory, which accurately models surface properties; (2) A large SVBRDF dataset for the Cook-Torrance model is publicly available \cite{DADDB18}.

We parameterize the environment lighting in the scene using the sky model proposed by \citet{HW12AAMFFSSDR}. Such a sky model simulates realistic and plausible environment lighting given only the sun direction $\omega_{s}$ and turbidity (cloudiness) $c$ as input. Hence, we can simulate large variations in the outdoor lighting with only four parameters.

We formally define the task of neural rendering as follows. Given the Cook-Torrance material parameters $m_{f}$ along with the incoming sun direction $\omega_{s}$ and turbidity $c$, the solution of the rendering equation is estimated by a convolutional neural network $\phi$ as:
\begin{equation}
    I(x) = \phi(x, m_{f}, \omega_{s}, c).
\end{equation}
We do not explicitly parameterize the geometry of our target scene since it remains constant across all renders. 

\begin{figure*}
    \centering
    \includegraphics[width=\linewidth]{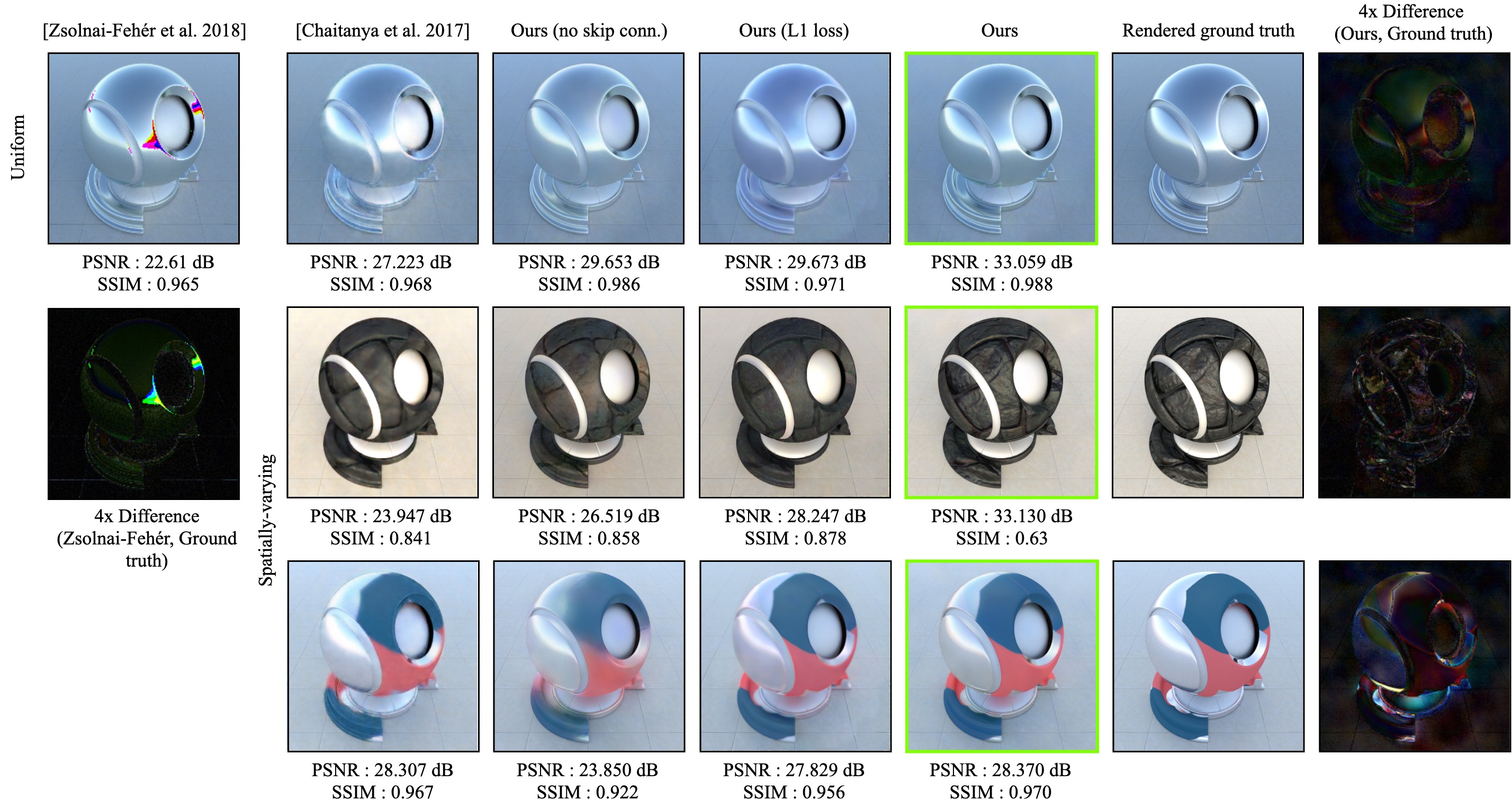}
    \caption{Average case and ablation study comparisons with \cite{DBLP:journals/tog/Zsolnai-FeherWW18} and \cite{Chaitanya:2017:IRM:3072959.3073601}. \cite{DBLP:journals/tog/Zsolnai-FeherWW18} has no results for spatially-varying materials, since they only handle uniform materials. Results are shown on a fixed sun direction.}
    \label{fig:comparison}
\end{figure*}

\subsection{Network architecture}
Figure \ref{fig:architecture} shows the overview of our proposed workflow. From input SVBRDF maps, we first construct their corresponding screen-space maps, by UV mapping each input map to the shaderball, and rasterizing the scene with that map as the base texture (Figure \ref{fig:architecture}(b)). The concatenation of these screen-space maps, along with the sun direction $\omega_{s}$ and turbidity $c$ forms the input to our network. Figure \ref{fig:architecture}(d) shows the rendered material output under different environment lighting conditions.

Our network architecture is inspired from the U-net-style autoencoder architecture \cite{DBLP:journals/corr/RonnebergerFB15}. The encoder takes the concatenation of 400x400 screen space maps of the material parameters: \textit{Diffuse}, \textit{Specular}, \textit{Roughness} and \textit{Normal}, and passes it through a series of convolutional layers, with stride 2 for downsampling. Each layer is followed by Batch Normalization and ReLU activation. We encode the 3D vector for the directional light using a separate, fully-connected encoder. Each fully-connected layer of this encoder is followed by \textit{Tanh} activation. The encoder expands the 3-dimensional vector to a 625-dimensional vector. We then reshape this 625 dimensional vector to a 25x25 dimensional feature map, and replicate it 128-times along the channel dimension, to get a 128x25x25 feature map. We append this feature map to the bottleneck layer of the Ray-Trace network. The decoder then deconvolves the concatenated feature map and encoder output. Each decoder layer is followed by Batch Normalization and ReLU activation. We use skip connections to recover the high frequency details in the rendering and improve convergence. The last layer of the decoder uses \textit{Sigmoid} activation and converts the 64 channel feature map to an output 3 channel target RGB image.

\subsection{Generating training data}
We generate a synthetic dataset of 50,000 material parameter maps and ground truth render pairs, containing equal number of spatially-varying and uniform parameter maps. We randomly choose 1000 images from the above dataset for the test set, and train our network on the remaining 49,000 images. For uniform maps, we randomly choose one value for all the four parameters (Diffuse, Specular, Roughness, Normal), and replicate it along the width and height (400x400) to get a uniform parameter map. We use SVBRDF textures from the dataset of \citet{DADDB18} for spatially-varying maps. For each material parameter map, we sample 5 random sun directions on the upper hemisphere with random turbidity value, and render the scene at 150 samples-per-pixel (spp). We use the cycles ray-tracing engine in Blender 3D \cite{blender3d}, in which we create the Cook-Torrance material for use on the shaderball. This dataset is available at \footnote{\url{https://aakashkt.github.io/neural-renderer-material-visualization.html}}.

\subsection{Training details}
The task of material visualization requires that the perceptual visual quality of the rendered images is impeccable. Cost functions based on Euclidean ($L_2$)  distance are known to be prone to blurring and pixel degradation. We therefore use a loss term which evaluates the perceptual quality of the rendered image, along with $L_1$ loss for training, inspired from \cite{DBLP:journals/corr/JohnsonAL16}.

Specifically, we use the feature reconstruction loss from a pre-trained VGG16 \cite{Simonyan2015VeryDC} network, which is given by :
\begin{equation}
    \mathcal{L}_{feat}^{j}(y', y) = \frac{1}{C_{j}H_{j}W_{j}}\left \| \phi_{j}(y') - \phi_{j}(y) \right \|_{2}^{2},
\end{equation}
where $\phi_{j}$ is the activation of the \textit{jth} convolutional layer with dimensions $C_{j}\times H_{j}\times W_{j}$ representing number of channels, width and height of the feature map, respectively. Here, $y$ denotes the predicted output and $y'$ is the ground truth. We use the \textit{relu\_3\_3} (\textit{j=relu\_3\_3}) feature representation in our experiments. Thus, our final composite loss is given by:
\begin{equation}
    \mathcal{L}_{train} = L_1 + \mathcal{L}_{feat}^{relu\_3\_3}.
\end{equation}

We train our network on an NVIDIA GTX 1080Ti, with a batch size of six using the Adam Optimizer (\textit{lr} = $10_{}^{-2}$, \textit{$\beta_{1}$} = 0.9, \textit{$\beta_{2}$} = 0.999). We initialize all weights using Glorot-initialization. The network is trained for 30 epochs and takes around 90 hours to train on our setup.

\begin{figure}
    \centering
    \includegraphics[width=\linewidth]{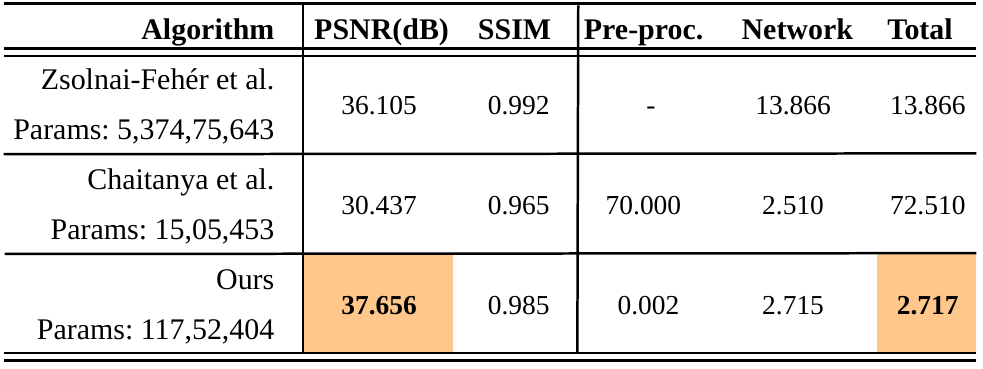}
    \caption*{Table 1: Quantitative and run-time comparisons (in milliseconds). The network of [Chaitanya et al. 2017] has a pre-process time for rendering the 2spp image, our network has a pre-process time for UV-mapping. Run-time values are evaluated on a workstation with 40 CPU cores and one NVIDIA GTX 1080Ti GPU.}
    \label{tab:quantative_comparison}
\end{figure}

\begin{figure*}[h]
    \centering
    \includegraphics[width=\linewidth]{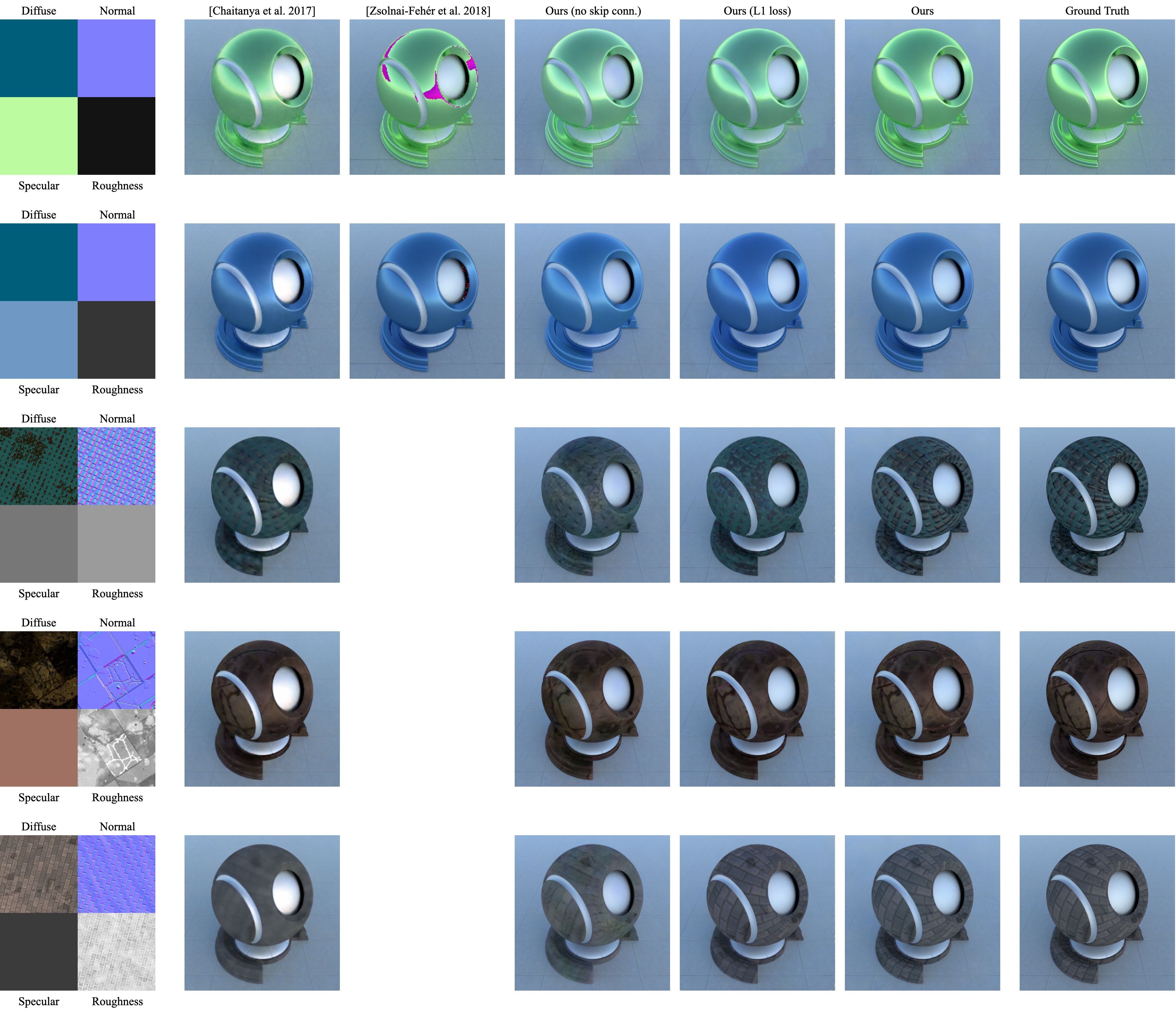}
    \caption{Comparisons with \cite{DBLP:journals/tog/Zsolnai-FeherWW18}, \cite{Chaitanya:2017:IRM:3072959.3073601} and ablations of our network. Results are shown for both uniform material parameter maps and spatially-varying material parameter maps. \cite{DBLP:journals/tog/Zsolnai-FeherWW18} has blank spots for spatially-varying materials, since they only handle uniform materials. Results are shown on a fixed sun direction.}
    \label{fig:more_comparison}
\end{figure*}

\section{Results and evaluation}
We compare our performance with two contemporary paradigms for neural rendering: (1) Neural rendering based on denoising \cite{Chaitanya:2017:IRM:3072959.3073601}; (2) Direct neural rendering \cite{DBLP:journals/tog/Zsolnai-FeherWW18}. We also justify our network design choices with an ablation study and conduct a user study for perceptual evaluation. Figures \ref{fig:more_comparison}, \ref{fig:more_results}, \ref{fig:light_1} and \ref{fig:light_2} show extensive results and comparisons.

\subsection{Comparison with denoising}
We show qualitative and run-time comparisons with the denoiser proposed by \citet{Chaitanya:2017:IRM:3072959.3073601}. We implement their network in PyTorch and train on our dataset with 2spp render input and 150spp ground truth, consisting of both uniform and spatially-varying materials. The denoiser fails to recover accurate details, which are essential for material visualization (Fig \ref{fig:comparison}). In terms of run-time, it is evident that first rendering a 2spp image and denoising it requires a lot more time than what is required by our network, even with the additional overhead of UV-mapping (Table 1).

\subsection{Comparison with direct neural rendering}
We also compare our results with the neural renderer proposed by \citet{DBLP:journals/tog/Zsolnai-FeherWW18}. We implement their network in PyTorch and train on our dataset of uniform material parameter maps. For a fair comparison, we compare results on a fixed sun direction. Our network produces better results than their network both in terms of render quality and PSNR values (Fig. \ref{fig:comparison}, \ref{fig:more_comparison} and Table 1). Since the number of parameters of our network differs from theirs by a factor of 10, the run-time of our network is also better.

\subsection{Ablation study}
We justify the impact of our composite loss function by training a standalone network using only the $L_1$ loss. Figure \ref{fig:comparison}, \ref{fig:more_comparison} demonstrates that $L_1$ loss alone is unable to capture fine details, especially of the normal map. We also justify the benefit of using skip connections in our network. Figure \ref{fig:comparison}, \ref{fig:more_comparison} shows this comparison. Without skip connections, several high-frequency details are lost (Fig. \ref{fig:comparison}, last two rows).

\subsection{Quantitative evaluation}
Table 1 and Fig. \ref{fig:comparison} show that quantitative metrics like PSNR and SSIM do not faithfully reflect the visual quality of results. Although the average PSNR value of \citet{DBLP:journals/tog/Zsolnai-FeherWW18} are comparable to ours, their result contains artifacts in near perfectly dark or white regions. Another point to note is the resultant PSNR values produced by the denoiser and by our network. The quantitative values are very close, even though the latter's visual quality is superior (Fig. \ref{fig:comparison}, last row).

\begin{figure}[h]
    \centering
    \includegraphics[width=\linewidth]{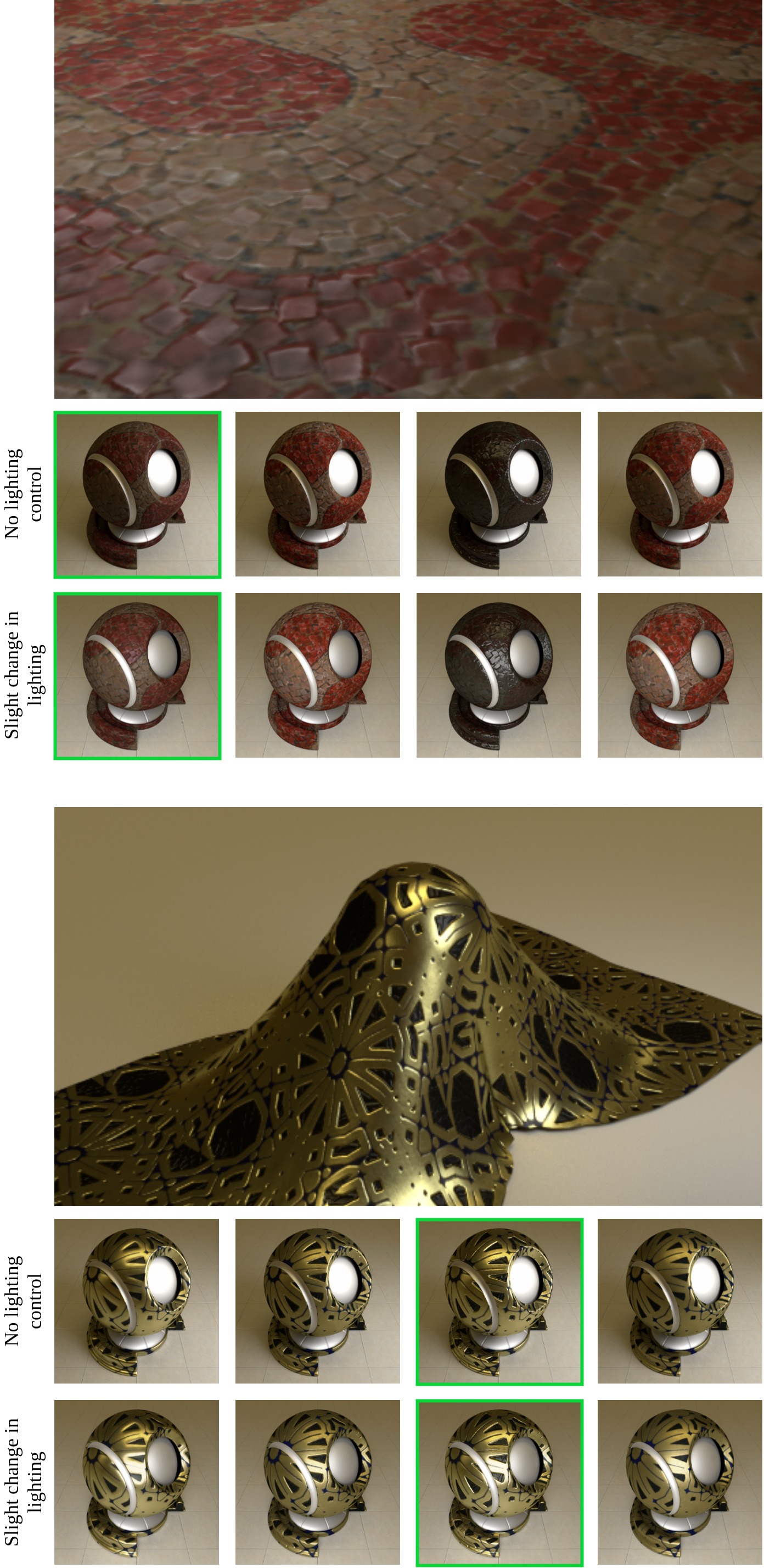}
    \caption{Two example questions and options presented to the users in our user study. The user was asked to identify the material in the scene. The second row of options are where lighting control was allowed. On viewing under certain lighting, the distinction between materials becomes clear. The correct option is highlighted in green.}
    \label{tab:user_study}
\end{figure}

\subsection{Qualitative user study}
We conduct an extensive user study consisting of 70 users to qualitatively validate the impact of flexible lighting for the task of material selection. From a rendered scene, we pick one dominant material from the scene and render four materials on the shaderball, with one of them being the correct material and others encoding slight variations. We ask the users to pick the correct materials in two independent experiments conducted with and without the freedom to view the shaderball under flexible lighting. Two example instances of the user study are shown in Figure \ref{tab:user_study}. We find that only 17.9\% of users are able to identify the correct material under fixed lighting conditions. This number increases to 49.3\% once the flexible lighting is made available. This clearly demonstrates the benefit of using flexible lighting in our renderings for material visualization.

\section{Conclusion}
In summary, we present a convolutional neural network for accurate rendering and visualization of both uniform and spatially-varying materials. We enable control over the environment lighting through our architecture, and verify its benefit through a qualitative user study. Comparison with denoising and neural rendering methods shows improved quantitative and qualitative results. We also release a large-scale dataset of uniform and spatially-varying material parameter-render pairs. In the future, we are interested in generalizing the network to arbitrary geometry.

\begin{acks}
We thank the reviewers of our SIGGRAPH Asia 2019 submission for their valuable comments and suggestions. \\
\\
\\
\\
\end{acks}

\begin{figure*}
    \centering
    \includegraphics[width=\linewidth]{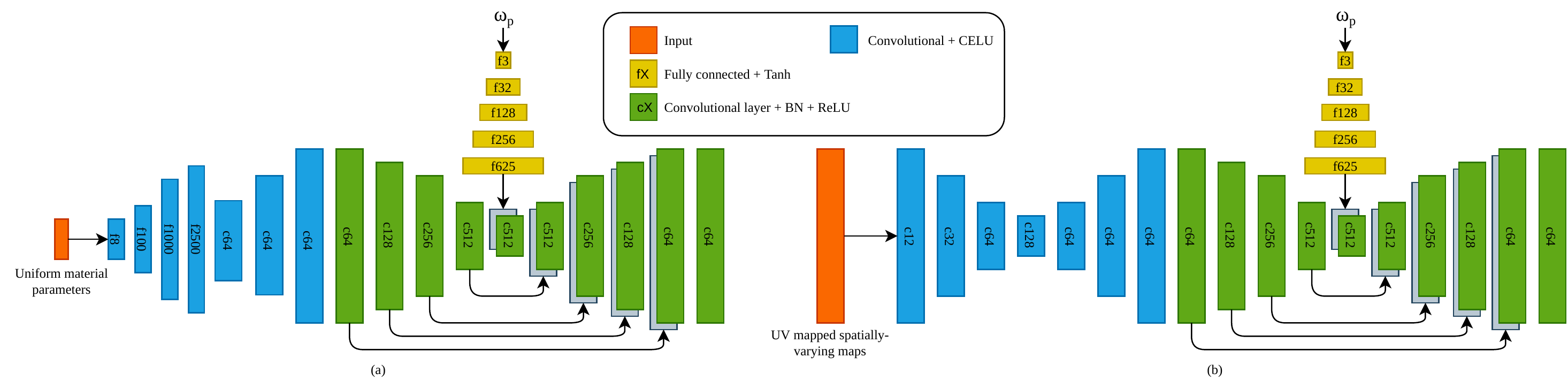}
    \caption{(a) Network to render a uniform materials on the shaderball. (b) Network to render spatially-varying materials on the shaderball. Each network was trained separately on uniform material dataset and spatially-varying material dataset, respectively.}
    \label{fig:m_svm_net}
\end{figure*}

\begin{figure*}
    \centering
    \includegraphics[width=0.7\linewidth]{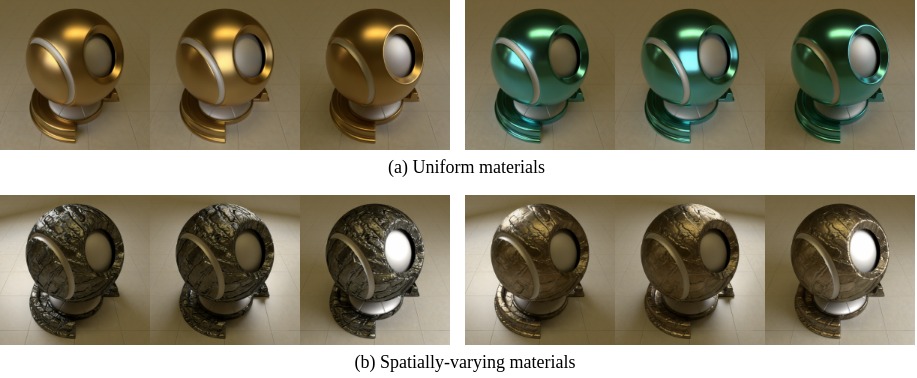}
    \caption{(a) Results of the network described in Figure 6(a) (Uniform materials) and (b) Results of the network described in Figure 6(b) (Spatially-varying materials), for different locations of the planar light source.}
    \label{fig:m_svm_net_results}
\end{figure*}

\appendix

\section{Preliminary experiments}
In this section, we describe various other experiments we conducted for material visualization. To improve the neural renderer proposed by \citet{DBLP:journals/tog/Zsolnai-FeherWW18} while also enabling flexible lighting, we used the network architectures shown in Figure \ref{fig:m_svm_net}. 

Figure \ref{fig:m_svm_net}(a) shows a network architecture which is an extension of \citet{DBLP:journals/tog/Zsolnai-FeherWW18}, which provided control over an area light source in the scene. We trained this network on a dataset of uniform materials using perceptual loss along with the $L_{1}$ loss, as described in this paper. Figure \ref{fig:m_svm_net_results}(a) shows some results from this network. We achieved better quantitative and qualitative results, on comparison with \cite{DBLP:journals/tog/Zsolnai-FeherWW18}, which further motivated the extension to handle spatially-varying materials.

Consequently, we used the network shown in \ref{fig:m_svm_net}(b) to handle spatially-varying materials. We provided a UV mapped material map as input to the network (Sect. 3.1), since spatially-varying materials can not be defined using singular values. We therefore used only convolutional layers (highlighted in blue), in place of fully connected + convolutional layers of Figure \ref{fig:m_svm_net}(a). We trained this network on a dataset of spatially-varying materials, using the training loss described in this paper. Results of this network are shown in Figure \ref{fig:m_svm_net_results}(b).

Both of the networks described above provided control over lighting through an area light source, whose location was restricted to a single ring on the upper hemisphere. Moreover, uniform materials are a special case of spatially-varying materials, which makes the two networks redundant. We thus propose a single network for handling both uniform and spatially-varying materials in this paper (Sect. 3.1). We also extend our network to handle a full sky model \cite{HW12AAMFFSSDR}, with sun locations defined at any point on the upper hemisphere. This was motivated by the great fidelity of both the preceding networks to handle arbitrary light locations on the ring of the upper hemisphere, although trained on randomly sampled light locations.

\begin{figure*}
    \centering
    \includegraphics[width=\linewidth]{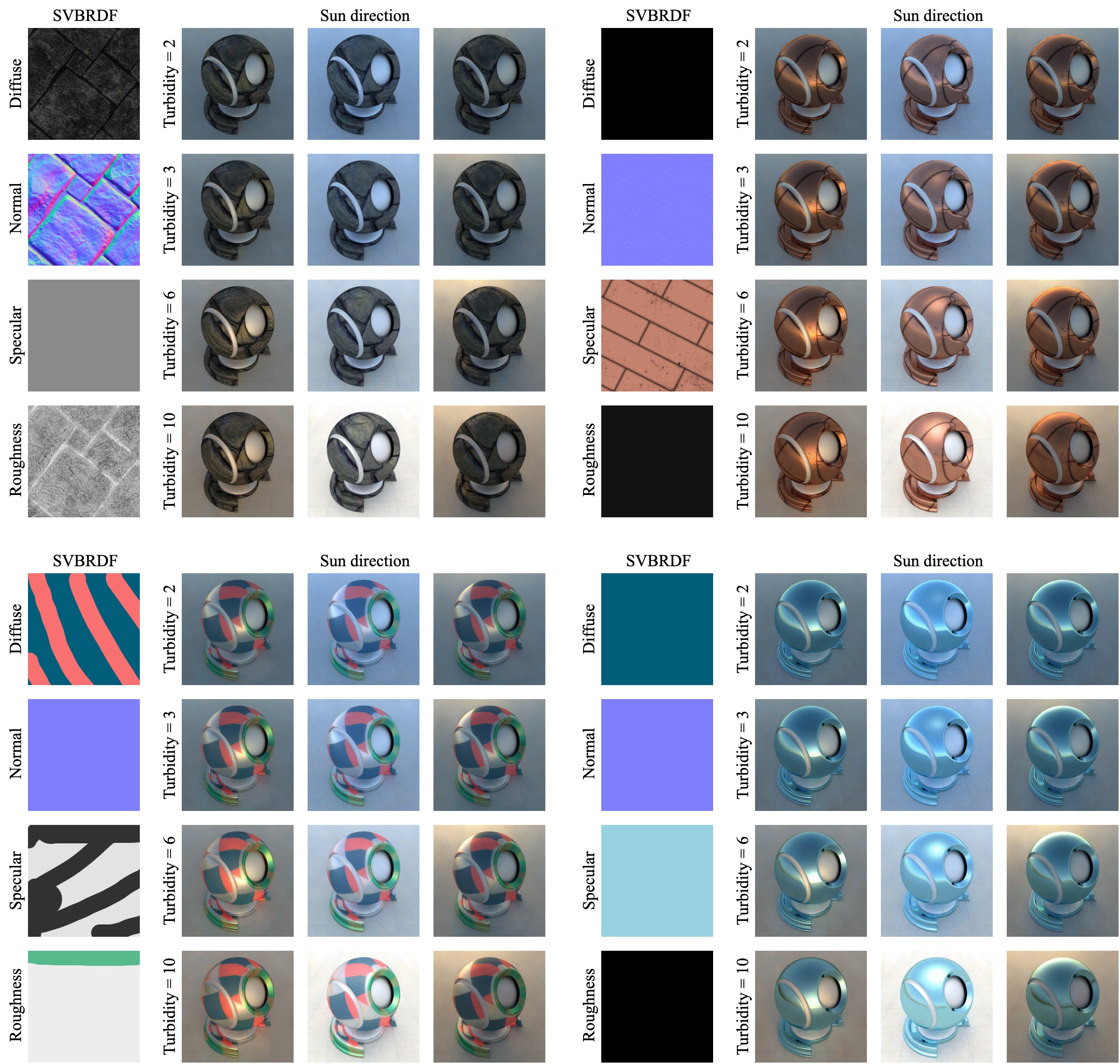}
    \caption{Shaderball visualizations under different environment lighting produced by our network corresponding to the input SVBRDF maps.}
    \label{fig:more_results}
\end{figure*}

\begin{figure*}
    \centering
    \includegraphics[width=\linewidth]{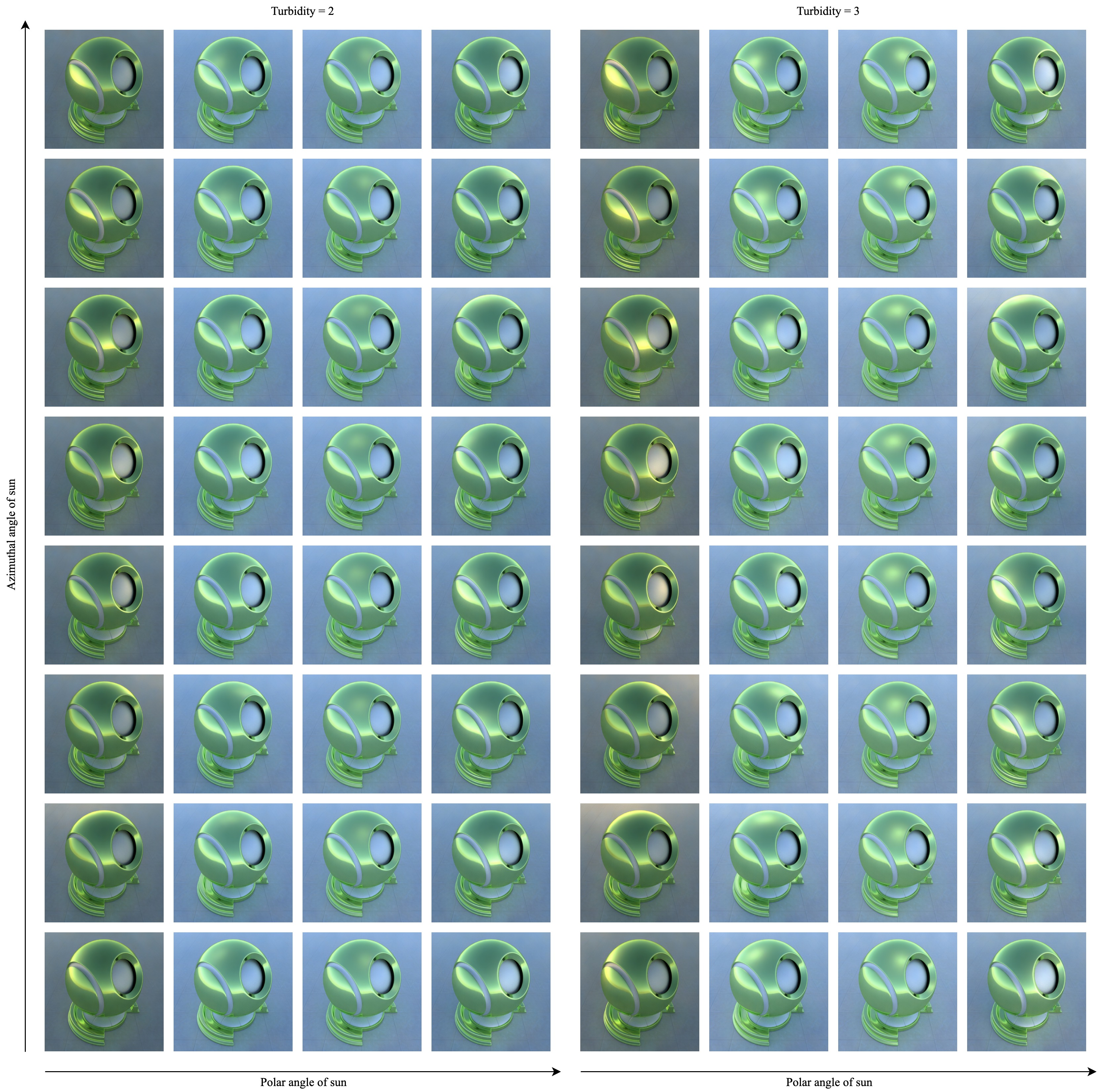}
    \caption{Visualization results for different sun directions and turbidity values of a specular uniform material.}
    \label{fig:light_1}
\end{figure*}

\begin{figure*}
    \centering
    \includegraphics[width=\linewidth]{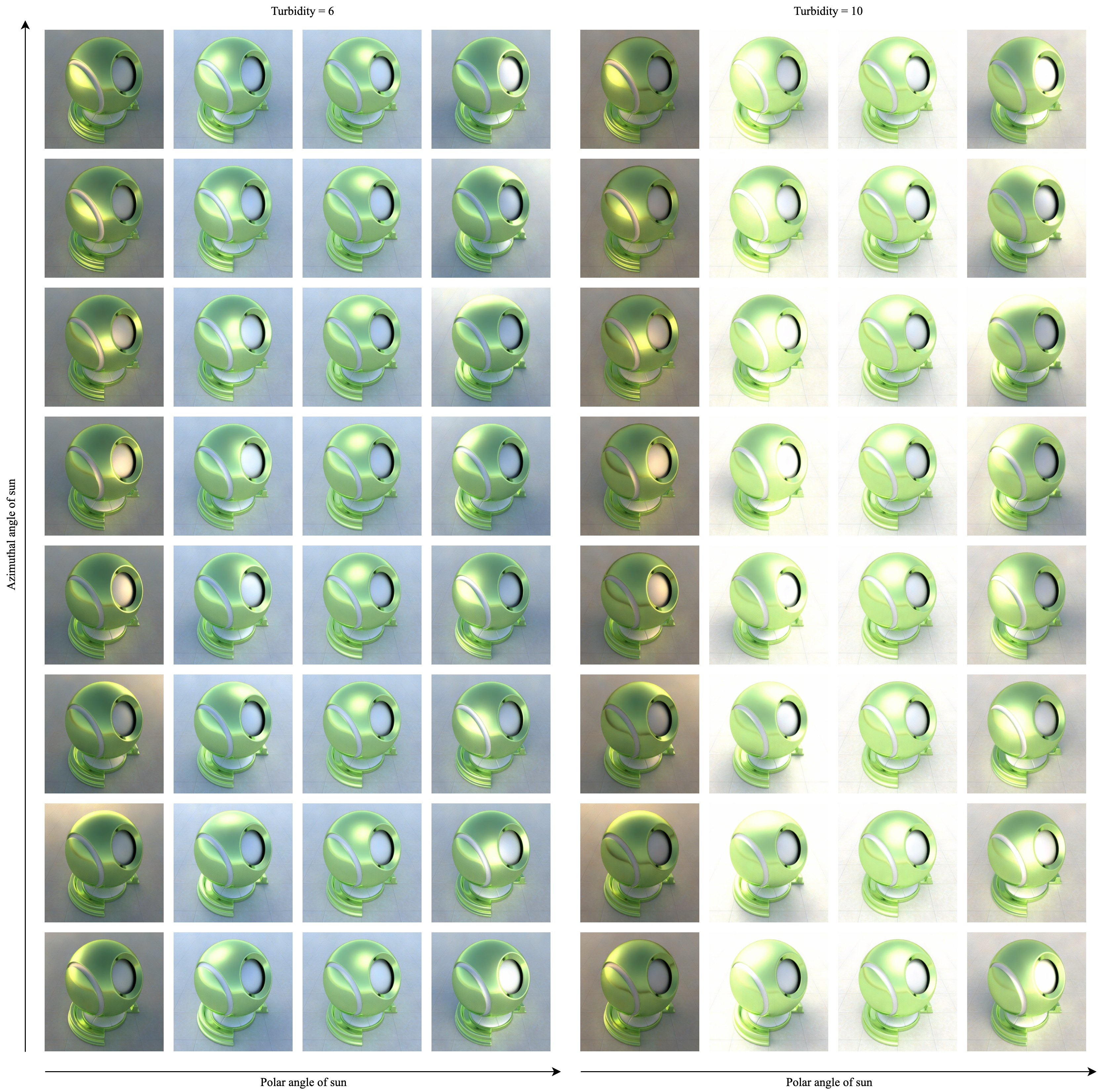}
    \caption{Visualization results for different sun directions and turbidity values of a specular uniform material.}
    \label{fig:light_2}
\end{figure*}

\bibliographystyle{ACM-Reference-Format}
\bibliography{main.bib}

\end{document}